\documentclass[superscriptaddress,preprint,amsmath,amssymb,aps,pra,]{revtex4-1}
\usepackage{graphicx}
\usepackage{bm}
\usepackage{color}
\usepackage{hyperref}
\hypersetup{colorlinks,citecolor=blue}
\begin{document}

\title{Raman sideband cooling of a $^{138}$Ba$^+$ ion using a Zeeman interval}

\author{Christopher M. Seck}
\affiliation{Department of Physics and Astronomy, Northwestern University, Evanston, Illinois 60208, USA}
\author{Mark G. Kokish}
\affiliation{Department of Physics and Astronomy, Northwestern University, Evanston, Illinois 60208, USA}
\author{Matthew R. Dietrich}
\affiliation{Department of Physics and Astronomy, Northwestern University, Evanston, Illinois 60208, USA}
\affiliation{Physics Division, Argonne National Laboratory, Argonne, Illinois 60439, USA}
\author{Brian C. Odom}
\email{b-odom@northwestern.edu}
\affiliation{Department of Physics and Astronomy, Northwestern University, Evanston, Illinois 60208, USA}

\date{\today}

\begin{abstract}
Motional ground state cooling and internal state preparation are important elements for quantum logic spectroscopy (QLS), a class of quantum information processing.  Since QLS does not require the high gate fidelities usually associated with quantum computation and quantum simulation, it is possible to make simplifying choices in ion species and quantum protocols at the expense of some fidelity.  Here, we report sideband cooling and motional state detection protocols for $^{138}$Ba$^+$ of sufficient fidelity for QLS without an extremely narrowband laser or the use of a species with hyperfine structure.  We use the two S$_{1/2}$ Zeeman sublevels of $^{138}$Ba$^+$ to Raman sideband cool a single ion to the motional ground state.  Because of the small Zeeman splitting, near-resonant Raman sideband cooling of $^{138}$Ba$^+$ requires only the Doppler cooling lasers and two additional AOMs.  Observing the near-resonant Raman optical pumping fluorescence, we estimate a final average motional quantum number $\bar{n}\approx0.17$.  We additionally employ a second, far-off-resonant laser driving Raman $\pi$-pulses between the two Zeeman sublevels to provide motional state detection for QLS and to confirm the sideband cooling efficiency, measuring a final $\bar{n} = 0.15(6)$.
\end{abstract}
\maketitle

\section{\label{sec:Intro}Introduction}

The qubits in atomic ion quantum computation and digital quantum simulation, subclasses of quantum information processing, require high-fidelity coherent operations between long-lived qubit states with error correction protocols typically requiring gate fidelities of $>$99\% \cite{Steane2003,Aliferis2007,Steane2007,Blatt2012}.  The atomic ion(s) are typically cooled to the motional ground states of the harmonic ion trap in order to facilitate quantum gate protocols and to minimize decoherence effects from residual thermal motion \cite{Cirac1995,Wineland1998,Brown2011}.  Quantum logic spectroscopy (QLS), a different subclass of quantum information processing, does not require the high gate fidelities needed for quantum computation; a lower gate fidelity only increases the experimental integration time required to achieve a given spectroscopic statistical uncertainty.  Most QLS-type protocols do, however, require or benefit from cooling at least one normal mode of the logic and spectroscopy ion pair to the motional ground state of the ion trap \cite{Schmidt2005,Rosenband2007,Lin2013,Wan2014,Wolf2016}.

Barium is one of the many Group II atomic ions that can be used as a coolant and logic ion for QLS.  Barium has a large mass, convenient for sympathetic cooling of heavy atomic or molecular ions of interest to precision measurement.  It also has the reddest Doppler cooling transition wavelength of the Group II ions, useful for avoiding unwanted laser coupling between molecular electronic states and also for integration with fiber optic technology \cite{Auchter2014}.

For high-fidelity coherent operations or sideband cooling of Ba$^+$, a 1.76 $\mu$m laser is sometimes used to drive the quadrupole S$_{1/2}$ $\rightarrow$ D$_{5/2}$ transition (D$_{5/2}$ lifetime $\sim35$~s \cite{Gurell2007}) \cite{Yu1994,Dietrich2010,Slodicka2012}.  Despite being a narrow quadrupole transition with a long lifetime, a high state transfer fidelity, beneficial for efficient sideband cooling, requires both a large Rabi frequency and a narrow linewidth laser to avoid both thermal and laser linewidth decoherence effects.  With the D$_{5/2}$ lifetime of $\sim35$ s compared with the same state lifetime of $\sim1.1$ s in Ca$^+$ \cite{Barton2000}, the requirements for the 1.76 $\mu$m laser source are then demanding.  Unfortunately, rapid adiabatic transfer techniques only slightly relax this large Rabi frequency and narrow linewidth requirement \cite{Noel2012}.

Rather than using a narrow quadrupole transition, one can also sideband cool using a Raman transition, tailored to be narrow by choices of detuning and Rabi frequencies \cite{Marzoli1994}, between two quantum states, e.g., electronic \cite{Chuah2013}, hyperfine \cite{Monroe1995,Deslauriers2004,Epstein2007,DeVoe2001}, or Zeeman levels \cite{Roos2000,BlytheThesis,WebsterThesis,Home2009,Lin2013NIST,Schindler2013}.  For Ba$^+$, both electronic \cite{Chuah2013} and hyperfine levels \cite{DeVoe2001} have been used for Raman sideband cooling.  Choice of a hyperfine interval for Raman sideband cooling has the advantage that the states can be chosen to be magnetically insensitive, allowing the same lasers to be used for cooling and high-fidelity coherent operations.  On the other hand, Raman sideband cooling on a Zeeman splitting, previously demonstrated \cite{Roos2000,BlytheThesis,WebsterThesis,Home2009,Lin2013NIST}, allows use of even isotopes, which are simpler to cool and manipulate due to their lack of hyperfine structure.  Additionally, ground state hyperfine intervals for some ions are too large for easy access by AOM (8 GHz for $^{137}$Ba$^+$), and would require e.g. either phase-locking of separate lasers or EOM modulation with optical filtering for sideband cooling.  For these reasons, the Zeeman splitting of $^{138}$Ba$^+$ is an attractive system for Raman sideband cooling for QLS, in which high quantum gate fidelities are not required.  Here, each of the two Raman beams is red-detuned from the single-photon resonance.  We make this choice to simplify the experimental apparatus and prevent ion heating during beam alignment.  The advantages of using blue-detuning \cite{Roos2000,Lin2013NIST} will be discussed in Sec.~\ref{sec:improvements}.

\section{\label{sec:Apparatus}Apparatus}

The experiment is conducted in a linear radio frequency trap of single-ion-scale previously described in Ref.~\cite{Lin2013}.  The ion trap parameters are $r_0$ = 1.26 mm, $z_0$ = 0.95 mm, and $\Omega_\text{RF} = 2\pi\times23.420$ MHz with $>$6.2 kV$_\text{PP}$ applied to two opposing rod electrodes and the other two held at ground.  The trap endcap rod electrodes are held at 200 V.  The observed secular frequencies are $\omega_z = 2\pi\times1.14$ MHz and $\omega_r = 2\pi\times2.3$ MHz.  The experimental vacuum system pressure is $<10^{-10}$ Torr.

A single $^{138}$Ba$^+$ ion is loaded into the trap via $(1+1^\prime)$ resonance-enhanced multi-photon ionization of neutral barium emanating from an oven under trap center.  A 791 nm external cavity diode laser (ECDL) driving the $^1$S$_0$ $\rightarrow$ $^3$P$_1$ intercombination line provides isotope-selective loading \cite{Steele2007} while a 310 nm UV LED (SET Inc. UVTOP-310TO39BL) provides a the second photon energetic enough to surpass the ionization threshold \cite{Wang2011}.  Ion lifetimes in the trap without Doppler cooling are $>$48 hours.  Micromotion compensation is performed using the RF-photon correlation method \cite{Berkeland1998}.  The optical arrangement and relevant levels/transitions in $^{138}$Ba$^+$ are depicted in Fig.~\ref{fig:LevelsTrap}.

\begin{figure*}[h]
\centering
\includegraphics[height=1.0\textwidth]{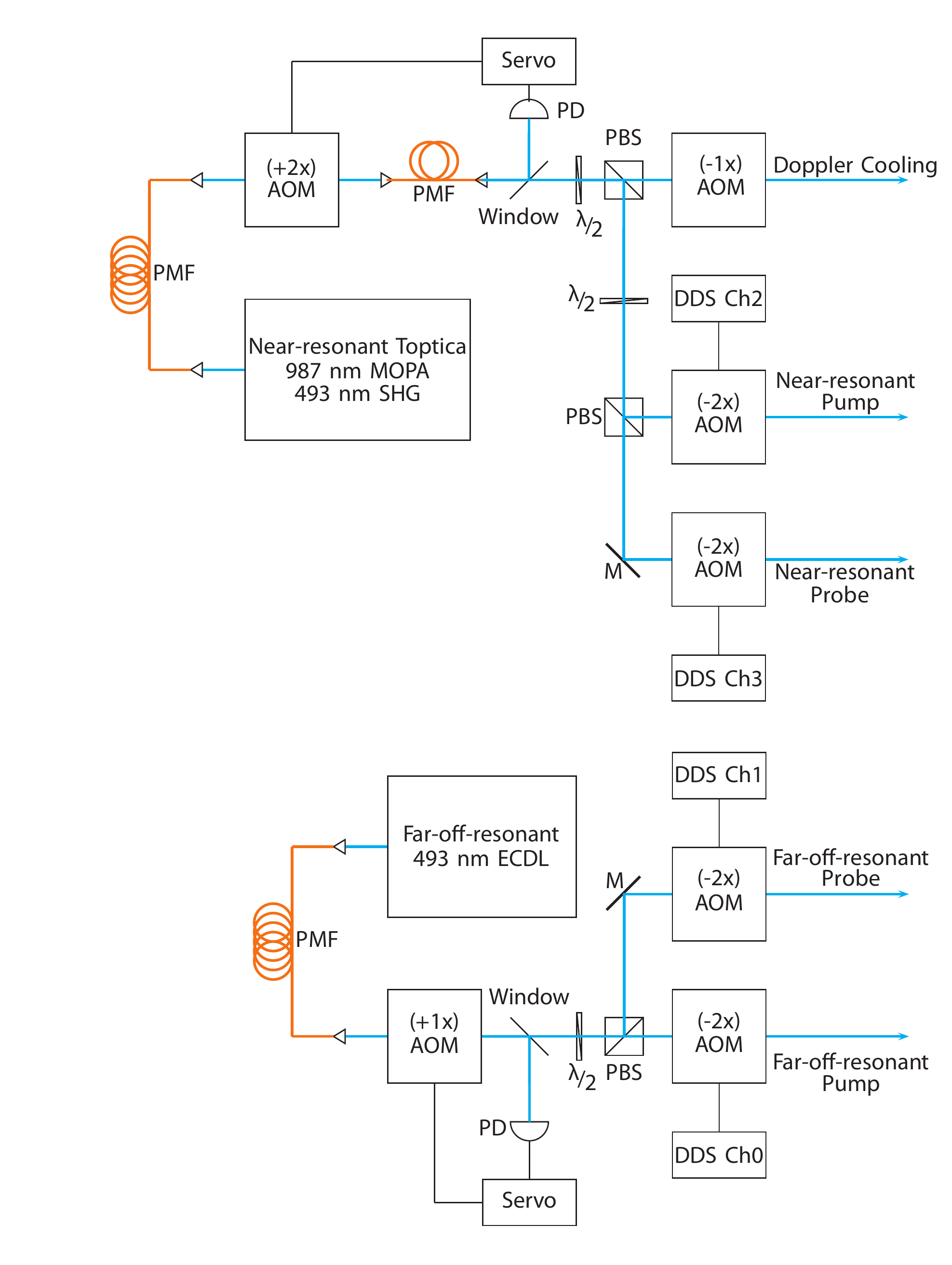}
\caption{\label{fig:493AOMs} AOM setup for the near- and far-off-resonant 493 nm beams.  A set of flip mirrors couples light from before the Doppler (-1x) AOM to the polarization-maintaining single-mode fiber (PMF) to the far-off-resonant (+1x) AOM.  This allows alignment of the far-off-resonant beam path using resonant light.  An amplified photodiode (PD) and servo circuit control the laser power to the double-pass Raman AOMs.  All AOM frequencies are $\sim$80 MHz.}
\end{figure*}

The 493 nm laser source for Doppler and near-resonant Raman sideband cooling is a Toptica 987 nm ECDL with tapered amplifier master oscillator power amplifier (MOPA) system with a second harmonic generation (SHG) cavity.  Fig.~\ref{fig:493AOMs} shows the AOM arrangement for the near-resonant Raman beams.  For the near-resonant Doppler and Raman beam source, the first AOM is in a +160 MHz double-pass configuration providing frequency shifting and power stabilization.  The second AOM in a -80 MHz single-pass configuration shifts the Doppler light to $\sim-\Gamma/2$ from the zero-field resonance and acts as the optical switch for the experiment.  The near-resonant Raman light is taken before the single-pass Doppler AOM, and is split into two AOMs each in -160 MHz double-pass configuration to generate the near-resonant Raman pump and probe beams $\sim$ -80 MHz from resonance.

A far-off-resonant Raman laser source is used as a tool to diagnose the near-resonant Raman sideband cooling.  This 493 nm laser source is an ECDL using a Nichia NDS1316 laser diode operating -59 GHz from the S$_{1/2}$ $\rightarrow$ P$_{1/2}$ transition.  Fig.~\ref{fig:493AOMs} shows the AOM arrangement for the far-off-resonant Raman beams.  The first AOM in a +80 MHz single-pass configuration provides power control.  The beam is then split to two AOMs each in the -160 MHz double-pass configuration to generate the far-off-resonant Raman pump and probe beams.

All four double-pass Raman AOMs are driven by a single AD9959 direct digital synthesizer (DDS) evaluation board with crystal oscillator clock.  This ensures phase coherence of all four AOM RF drive signals and permits computer control of the output frequencies and amplitudes over USB.  Each DDS channel is amplified by a preamplifier before passing through two TTL RF switches in series for $>$100 dB of extinction.  Additionally, a voltage-controlled-oscillator (VCO), also using two TTL RF switches in series, provides an $\sim$80 MHz drive for the shelving and deshelving laser sources.  All signals are then amplified to a maximum of 30 dBm.  All other AOM drives are commercial units.

The 650 nm Doppler repump laser is a Toptica DL100 ECDL.  The 614 nm deshelving laser source is the output of an AdvR Inc. SHG module pumped by an ECDL using an Innolume GC-1220-110-TO-200-B curved stripe gain chip.  The 455 nm shelve laser is a free-running Nichia NDB4216E laser diode.  All laser systems unless otherwise specified are lab-built including temperature and current controllers.  All ECDLs are locked by a High Finesse WSU10 wavelength meter with a lock bandwidth of $\sim$1 Hz.  All experimental control and timings are performed by LabVIEW and/or Python using the National Instruments NI-DAQmx platform.

\section{\label{sec:NRRaman}Near-Resonant Raman Sideband Cooling}

\begin{figure*}[h]
\centering
\includegraphics[width=1.0\textwidth]{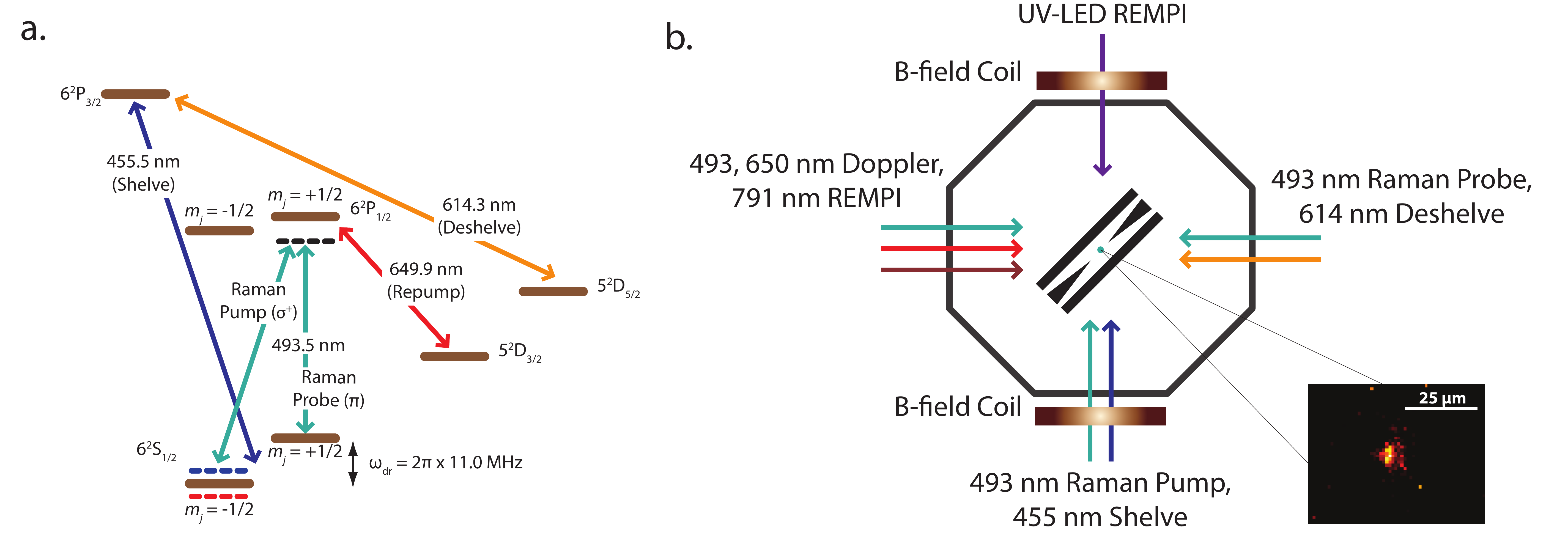}
\caption{\label{fig:LevelsTrap} (a) Relevant transitions in $^{138}$Ba$^+$.  All laser sources except the 455 nm free-running laser diode are ECDL systems locked by wavemeter.  Each laser source is shuttered via AOM and mechanical shutter with only the former used when fast timing is required. (b) Apparatus and beam geometry.  Note that the magnetic field is parallel to the $\sigma^+$-polarized Raman pump beam and perpendicular to the $\pi$-polarized Raman probe beam and that $\Delta\vec{k}$ lies along the trap $z$-axis.  Inset: an intensified CCD image of a single fluorescing $^{138}$Ba$^+$ in the ion trap.}
\end{figure*}

\begin{figure*}[h]
\centering
\includegraphics[width=1.0\textwidth]{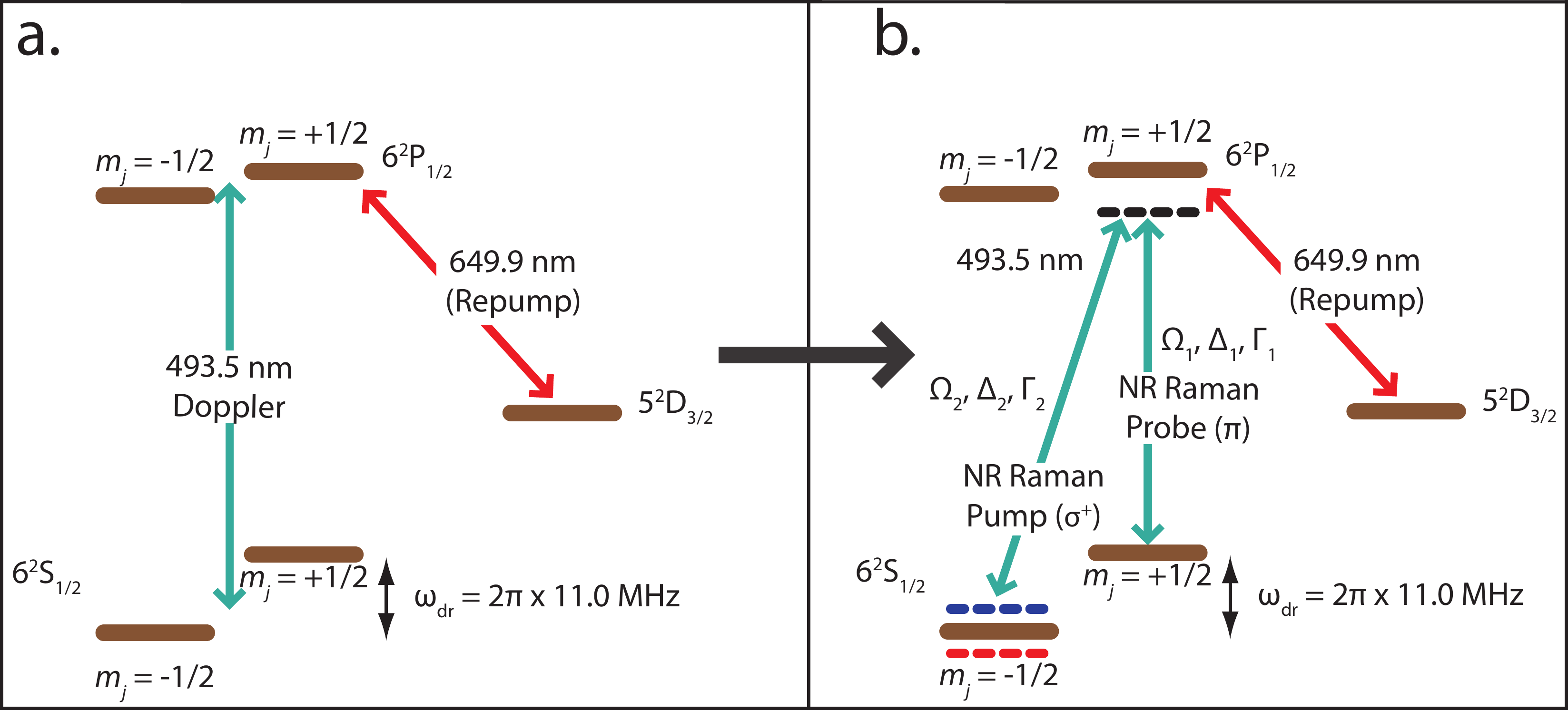}
\caption{\label{fig:CWRamanTiming}Stages of $^{138}$Ba$^+$ cooling.  Solid lines show applied laser drives.  Horizontal dashed lines show the red and blue motional sidebands of the driven Raman transition when relevant and the virtual state of the Raman transition (black), detuned by $\approx80$ MHz.  (a) Doppler cooling lasts 20 ms.  (b) Near-resonant Raman sideband cooling lasts either 1 or 10 ms.  The PMT is gated with the near-resonant Raman beams.}
\end{figure*}

The near-resonant Raman sideband cooling laboratory setup, AOM setup, and process are shown in Figs.~\ref{fig:LevelsTrap}, \ref{fig:493AOMs}, and \ref{fig:CWRamanTiming}, respectively.  The near-resonant Raman pump (probe) beams enter the trap parallel (perpendicular) to an applied magnetic field of 3.919(7) Gauss, which splits the S$_{1/2}$ ground state Zeeman sublevels by 10.97(2) MHz.  The magnetic field, oriented 45$^\circ$ from the \textit{z}-axis of the trap, is aligned to the $\sigma^+$-polarized near-resonant Raman pump beam using three additional coils, minimizing the spontanous emission rate from this beam.  Entering the trap 90$^\circ$ from the near-resonant pump beam, the $\pi$-polarized Raman probe beam provides a maximum value of $\Delta\vec{k}$ along the trap $z$-axis with our polarizations and geometry.

\begin{figure*}[h]
\includegraphics[width=1.0\textwidth]{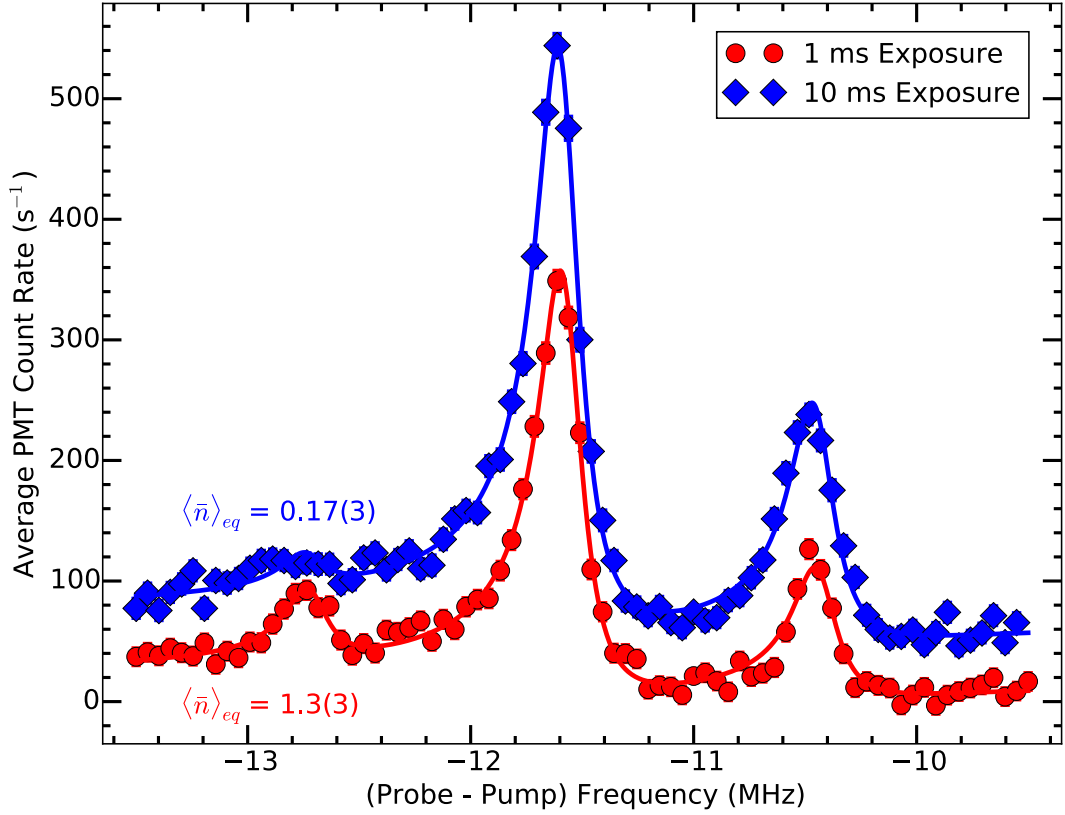}
\caption{\label{fig:CWRaman} Near-resonant Raman 493 nm fluorescence spectrum with red and blue motional sidebands resolved.  The red cirlces (blue diamonds) are data corresponding to PMT exposure for the first 1 (10) ms of near-resonant Raman exposure.  Solid lines are fits that can be used to extract equilibrium-equivalent occupation numbers, $\langle\bar{n}\rangle_{eq}$, discussed in Sec.~\ref{sec:NRRaman}.  Error bars are given by the standard deviation of the mean of each data point, and are comparable to the marker size.}
\end{figure*}

The two Zeeman S$_{1/2}$ sublevels serve as our quantum states for sideband cooling and motional state detection.  For Raman sideband cooling (Fig.~\ref{fig:CWRamanTiming}), the $\sigma^+$-polarized Raman pump beam with Rabi frequency $\Omega_\text{pump} = 2\pi\times14.9(3)$ MHz optically pumps the ion to the $m_j$ = +$\frac{1}{2}$ state via off-resonant spontaneous scattering and creates a light shift of 620(20) kHz.  With the near-resonant Raman probe beam sufficiently weak, the width of the carrier transition (those preserving the axial motional state $n$) decreases below the axial secular frequency, and the the axial secular sidebands are resolved.  In this regime, the near-resonant Raman probe beam with Rabi frequency $\Omega_\text{probe} = 2\pi\times1.07(2)$ MHz provides an effective Raman Rabi frequency of $\Omega_\text{eff} = 2\pi\times89(2)$ kHz.  When the frequency difference between the two Raman beams is near the light-shfited 2-photon Raman bright resonance or on either the red sideband (RSB; decrease $n$ by 1 motional quanta) or blue sideband (BSB; increase $n$ by 1 motional quanta), Raman transitions occur from the $m_j = +\frac{1}{2}$ state to the $m_j= -\frac{1}{2}$ state.  The ion is then quickly optically pumped to the $m_j = +\frac{1}{2}$ state by the $\sigma^+$-polarized near-resonant Raman pump beam.  With the 650 nm D$_{3/2}$ $\rightarrow$ P$_{1/2}$ repump laser on, this process results in the emission of 493 nm photons at a low rate similar to that of the optical pumping, which are collected on a PMT (Fig.~\ref{fig:CWRaman}).

The photon collection rate as a function of 2-photon detuning is shown in Fig.~\ref{fig:CWRaman} for two PMT exposure times with the near-resonant Raman pump frequency fixed while the near-resonant Raman probe frequency is scanned over the Raman resonance.  All data collection was performed with frequencies randomized and interleaved with a total exposure time of 20 s per data point.  In 1 ms of exposure, the RSB is well resolved.  In 10 ms of exposure, the RSB amplitude averages down to a low value since the ion cools to the ground state.  At this detuning and elsewhere red of the carrier, the carrier transition is still excited at a low rate, contributing a constant photon background.  A Fano-like fit function in the strong pump, weak probe limit \cite{Lounis1992,Stalgies1998,Morigi2000,McDonnell2004} is used with the subscript 1 (2) corresponding to the near-resonant Raman probe (pump) beam.  $\rho_{33}$, the equilibrium population in the excited P$_{1/2}$ state due to Raman transitions and optical pumping as a function of laser detuning $\omega$, is given by
\begin{equation}
\rho_{33}\left(\omega\right)\simeq\frac{\Omega_\text{eff}^2 \left(\frac{\omega}{\Delta^\prime}\right)^2\left(\frac{R}{4\Gamma_1}\right)}{\left(\omega-\Delta^\prime\right)^2+\frac{R^2}{4}+\frac{\Omega_\text{eff}^2\Gamma}{2\Gamma_1}}
\end{equation}
where
\begin{equation}
\Omega_\text{eff}=\frac{\Omega_1\Omega_2}{2\Delta_1}
\end{equation}
is the effective Raman Rabi frequency for oscillations on the 2-photon Raman resonance,
\begin{equation}
\Delta^\prime = \frac{\Omega_2^2}{4\Delta_1}
\end{equation}
is the light shift of the 2-photon bright resonance,
\begin{equation}
R=\frac{\Omega_2^2}{4\Delta_1^2}\Gamma
\end{equation}
is the off-resonant scattering rate on the strongly driven Raman pump transition, $\Gamma=\Gamma_1+\Gamma_2=\frac{1}{3}\Gamma+\frac{2}{3}\Gamma$ is the S$_{1/2}$ $\rightarrow$ P$_{1/2}$ natural linewidth and branching ratios for the Raman probe and pump transitions, and
\begin{equation}
\Delta_1=\Delta_2+\delta
\end{equation}
is the detuning of the near-resonant Raman probe beam from the S$_{1/2}$ $\rightarrow$ P$_{1/2}$ transition.  For our experiment, $\Delta_2=2\pi\times-79$~MHz so the 80-MHz-center-frequency near-resonant Raman probe AOM operates at its peak efficiency near the carrier bright resonance.  The full fit function for the near-resonant Raman spectrum is then the summation of $\rho_{33}$ for the carrier and each motional sideband with independent amplitudes:
\begin{equation}
A_\text{C}\rho_{33}\left(\delta-\omega_\text{dr}\right)+A_\text{RSB}\rho_{33}\left(\delta-\omega_\text{dr}+\omega_z\right)+A_\text{BSB}\rho_{33}\left(\delta-\omega_\text{dr}-\omega_z\right)
\end{equation}
where $A_\text{C}$ is the carrier amplitude, $A_\text{RSB}$ ($A_\text{BSB}$) is the red (blue) sideband amplitude, $\omega_\text{dr}$ is the location of the dark resonance (equal to the Zeeman splitting of the S$_{1/2}$ state), and $\omega_z$ is the axial secular frequency (both fit parameters).  Although all backgrounds are subtracted from the data, a constant background term is also included to account for any small background due to changes in the 493 nm light scatter over the data collection time.  The additional fit parameters are $\Omega_1$, $\Omega_2$, and $\omega_\text{dr}$.  Average fit values from each data set are given above.  Additionally, the fit provides the location of the RSB, which should correspond to the optimal Raman sideband cooling frequency for our system, which is verified in Sec.~\ref{sec:FORRaman}.

Although the data presented in Fig.~\ref{fig:CWRaman} represent time averages over timescales where cooling is occurring, it is instructive to extract from the sideband amplitudes a value $\langle\bar{n}\rangle_{eq}$ \cite{Leibfried2003} describing an equilibrium occupation number which would lead to these spectra.  With 1 ms of cooling, this value is $\langle\bar{n}\rangle_{eq} = 1.3(3)$, and with 10 ms of cooling, this value is $\langle\bar{n}\rangle_{eq} = 0.17(3)$, which is in agreement with our $\bar{n}$ measurement described in Sec.~\ref{sec:FORRaman}.  This suggests that the system reaches equilibrium in $\lesssim$10 ms.  Using this detection method with long exposure times, a good estimate of $\bar{n}$ can be made, and it is important to note that no additional laser systems are necessary to measure $\bar{n}$.  However, we use an additional, far-off-resonant laser system to conduct temperature diagnostics to confirm our final $\bar{n}$ value and verify the fitted RSB location provides the best cooling efficiency.

\section{\label{sec:FORRaman}Far-off-resonant Raman Diagnostics}

The near-resonant Raman sideband technique discussed above allows for an estimation of $\bar{n}$.  However, since the spontaneous emission rate is too high, it does not provide a mechanism to drive the $\pi$-pulses needed for single-shot measurement of $n$, as would be needed for QLS.  Far-off-resonant Raman excitation, achieved in our experiment using a single dedicated ECDL, allows the coherent Rabi flopping required for measurements of $\bar{n}$.

The far-off-resonant Raman laser source is detuned -59 GHz from the S$_{1/2}$ $\rightarrow$ P$_{1/2}$ transition and is used to coherently transfer population between the $m_j=+\frac{1}{2}$ and $m_j=-\frac{1}{2}$ Zeeman sublevels.  After the far-off-resonant Raman pulse, the population in each $m_j$ state must be measured.  This is performed using a shelving technique, common for Group II ions, in a spin-dependent method previously demonstrated in Ba$^+$ \cite{Koerber2003}.  The experimental procedure, preceeded by steps (a) and (b) in Fig.~\ref{fig:CWRamanTiming}, is outlined in Fig.~\ref{fig:FORRamanTiming} and proceeds as follows: (c) the ion is optically pumped to the $m_j=+\frac{1}{2}$ state; (d) the far-off-resonant Raman pulse is applied to transfer population, conditional on the motional state; (e) the near-resonant Raman pump "protects" population driven to the $m_j=-\frac{1}{2}$ state by transfering it to the D$_{3/2}$ manifold; (f) all remaining population in S$_{1/2}$ is shelved to the long-lived D$_{5/2}$ manifold, (g) the Doppler lasers detect whether the ion is in the D$_{3/2}$ protected state (ion fluoresces) or the D$_{5/2}$ shelved state (ion dark); and (h) any shelved ions are deshelved and the ion is Doppler cooled.

The net result of the above procedure is that a successful (unsuccessful) far-off-resonant Raman transition yields a bright (dark) ion when exposed to Doppler cooling light.  There are two critical times involved in the procedure above, the optical pumping time and the protect time.  Three contrast-reduction mechanisms occur in this detection scheme: (1) an imperfect $\pi$-pulse on the motional sidebands, (2) an imperfect protect stage (discussed in App.~\ref{PTAppendix}), and (3) an imperfect shelving stage.  For the latter, the branching ratio from P$_{3/2}$ $\rightarrow$ D$_{5/2}$ is only 88\% with the remaining 12\% branching to the protected D$_{3/2}$ state.

\begin{figure*}[h]
\includegraphics[width=1.0\textwidth]{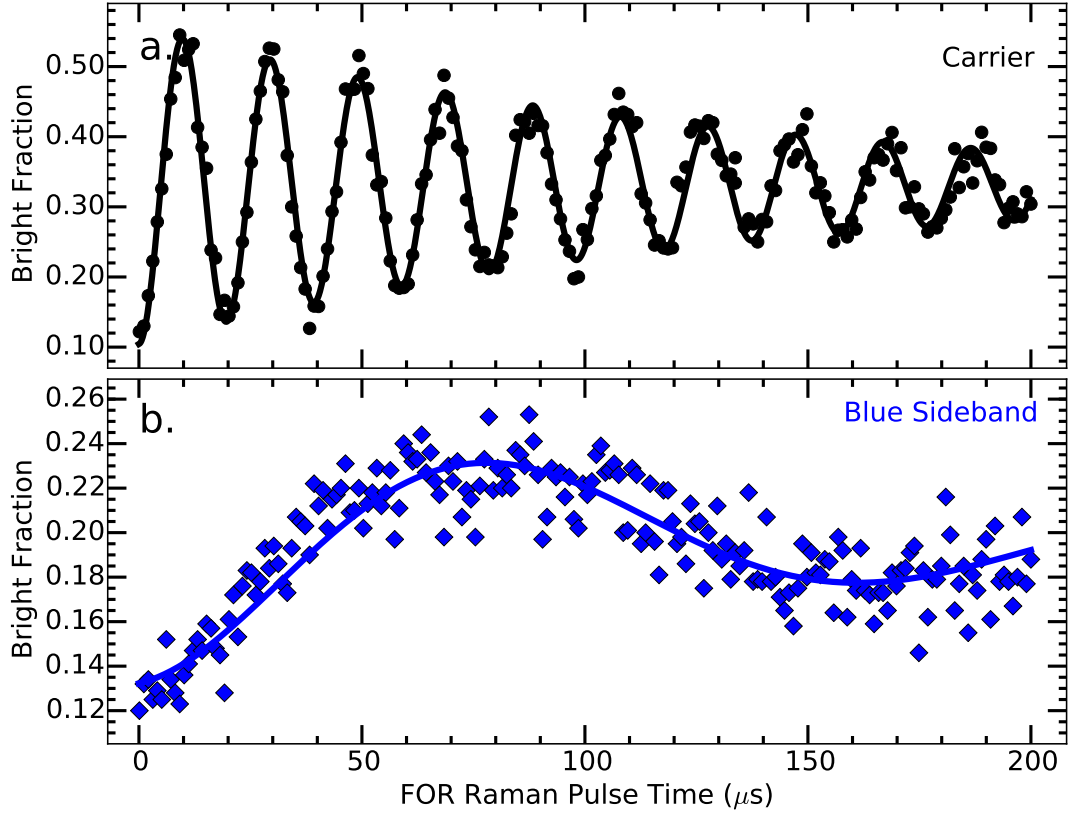}
\caption{\label{fig:FORTime} Far-off-resonant Raman excitation Rabi oscillations on (a) the carrier and (b) the BSB.  Multiple Rabi oscillations are observed on the carrier with $\Omega_\text{eff, carrier} = 2\pi\times50.92(3)$ kHz with a coherence time of 130(5) $\mu$s.  An efficient $\pi$-pulse on the BSB cannot be performed due to the coherence time of 134(11) $\mu$s being shorter than the $\pi$-pulse time with $\Omega_\text{eff, BSB} = 2\pi\times6.02(8)$ kHz.  Data points in each data set were taken randomly and interleaved.}
\end{figure*}

Using the optimal optical pumping and protect times described in App.~\ref{OPTAppendix} and \ref{PTAppendix}, respectively, the far-off-resonant pulse time was swept while observing the bright ion fraction on the carrier as shown in Fig.~\ref{fig:FORTime}a.  From the fit to the multiple Rabi oscillations, the far-off-resonant $\Omega_\text{eff, carrier} = 2\pi\times50.92(3)$ kHz, which is fast compared with the extracted decoherence rate of $2\pi\times7.7(3)$ kHz.  Raman laser intensity fluctuations, the residual thermal spread in the two radial motional states \cite{Brown2011}, and non-zero $\bar{n}$ value in the cooled axial motional state (Debye-Waller factors \cite{Wineland1998}) are all possible sources of the observed decoherence.  Consistent with the latter possibility, we observe the decoherence rate increase by a factor of 4 when not applying Raman sideband cooling.

Fig.~\ref{fig:FORTime}b shows the same experiment performed on the BSB.  Due to the weak coupling to the motional mode, the BSB Rabi frequency is an order of magnitude lower, $\Omega_\text{eff, BSB} = 2\pi\times6.02(8)$ kHz, which is simlar to the decoherence rate of $2\pi\times7.5(2)$ kHz.  Thus, the amplitude of the Rabi oscillation at the $\pi$-pulse time is significantly decreased.  Assuming that the ion is in the motional ground state for the Rabi oscillation experiments, the ratio of Rabi frequencies yields the Lamb-Dicke parameter $\eta=0.118(2)$, comparable to the theoretical 0.102 for $^{138}$Ba$^+$ using our ion trap parameters.

To measure $\bar{n}$ using the far-off-resonant Raman technique, the far-off-resonant 2-photon detuning was swept while driving $\pi$-pulses on the carrier, RSB, and BSB as shown in Fig.~\ref{fig:FORTemp}.  Five frequency ranges were collected: the full carrier, RSB, BSB, and a region nearby each sideband to constrain the background.  The data were fit to three Lorentzian lineshapes with the sidebands constrained to share the same width, $\omega_z$ displacement from the carrier, and a constant background term.  When no near-resonant Raman sideband cooling is applied, equal-amplitude red and blue sidebands are observed, indicating $\bar{n}\gg1$.  With 10 ms of Raman sideband cooling, the RSB amplitude decreases to a near-zero value of 0.017(6), and comparing the sideband amplitudes yields $\bar{n}$ = 0.15(6), consistent with the near-resonant Raman PMT result of $\langle\bar{n}\rangle_{eq} = 0.17(3)$ with 10 ms of exposure.

\begin{figure*}[h]
\includegraphics[width=1.0\textwidth]{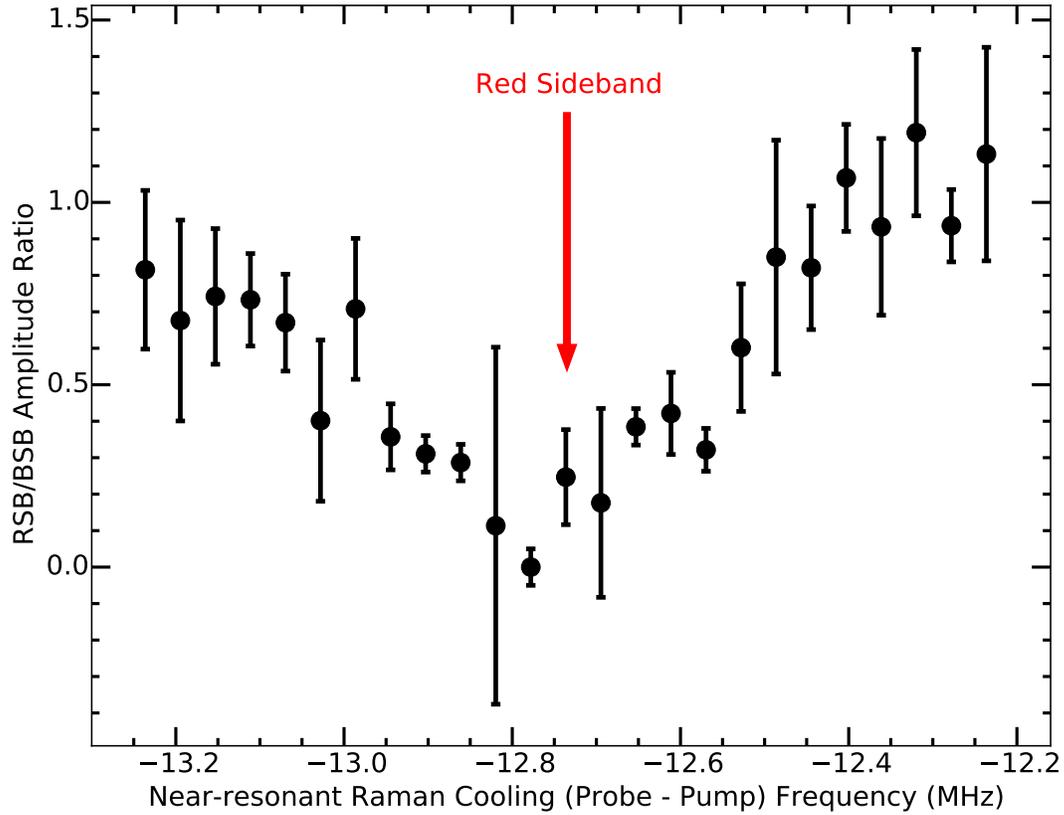}
\caption{\label{fig:CWTemp} Near-resonant Raman frequency scan around the RSB while measuring the RSB/BSB amplitude ratio (ion temperature) with 10 ms of sideband cooling.  The minimum amplitude ratio is reached while sitting on the peak of the red sideband measured by the near-resonant Raman fluorescence.  Data points were taken randomly and interleaved.  Error bars are given by the propagation of the the least squares curve fit errors for the red and blue sideband amplitudes.}
\end{figure*}

To verify that the lowest axial motional temperature is achieved while performing near-resonant Raman sideband cooling on the RSB, the near-resonant 2-photon detuning was swept while measuring the final axial RSB to BSB amplitude ratio after 10 ms of cooling as shown in Fig.~\ref{fig:CWTemp}.  This data was collected using a relatively fast 4-point measurement in which the measured frequencies were the RSB and BSB peaks and corresponding nearby background points.  The corresponding background point value was subtracted from the sideband peak value before the sideband amplitude ratio was calculated.  From the data, the lowest temperature is achieved close to the peak of the near-resonant RSB location, as expected.

\section{Potential Apparatus Improvements\label{sec:improvements}}

Although this work demonstrates sufficient experimental motional state preparation and detection contrast for QLS, a few simple improvements will decrease the final $\bar{n}$ value, stabilize the Zeeman splitting, remedy the low sideband Rabi oscillation amplitude, and increase the far-off-resonant detection contrast.

The ion could be cooled to lower $\bar{n}$ using electromagnetically induced transparency (EIT) cooling \cite{Morigi2000}, a more sophisticated version of Raman sideband cooling.  In this method, both the near-resonant Raman pump and probe beams are \textit{blue-detuned} from the S$_{1/2}$ $\rightarrow$ P$_{1/2}$ transition.  In this system, the RSB and BSB occur on the opposite side of the carrier as compared with Fig.~\ref{fig:CWRaman}.  By adjusing the strong near-resonant Raman pump Rabi frequency and hence imparted light shift, the RSB can be placed at the dark resonance of the carrier.  This prevents off-resonant excitation of the carrier, which is one limiting mechanism for red-detuned Raman sideband cooling \cite{Roos2000,Lin2013NIST}.

\begin{figure*}[h]
\includegraphics[width=1.0\textwidth]{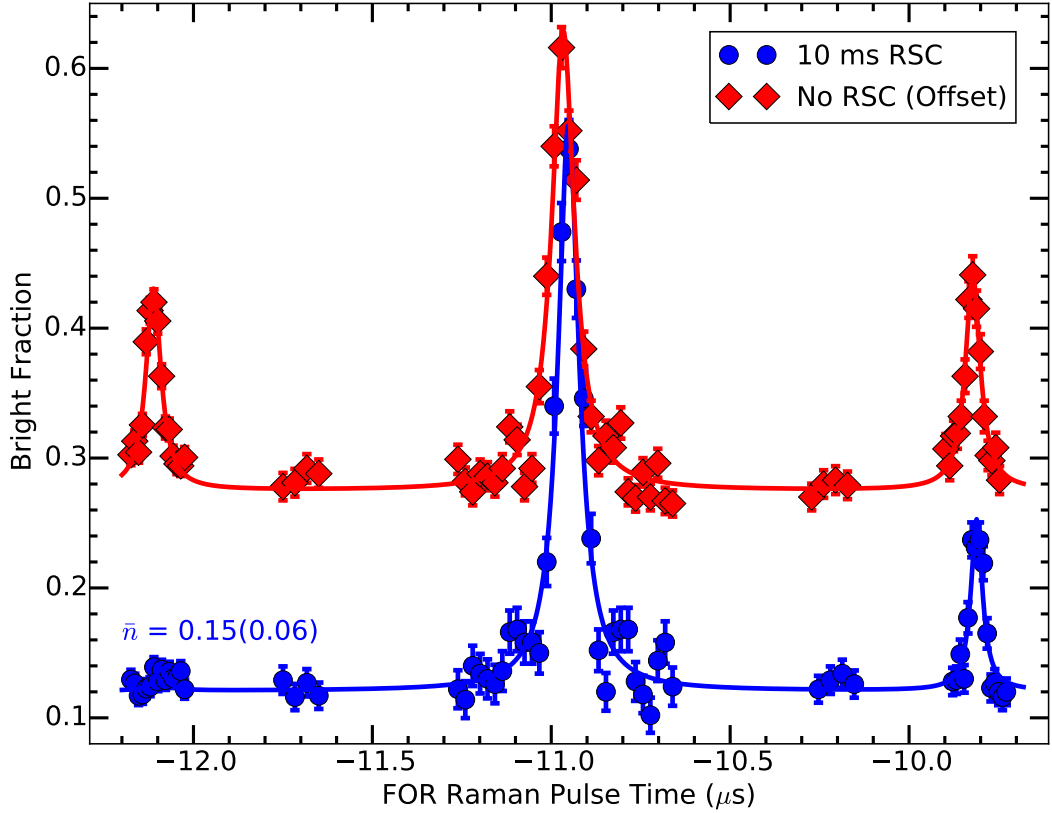}
\caption{\label{fig:FORTemp} Far-off-resonant Raman frequency spectrum.  A $\pi$-pulse is used on the carrier.  The longer sideband $\pi$-pulse is used on the RSB and BSB as well as the background points that help the fit establish the background shelving level.  Near equal blue and red sideband amplitudes are observed if no near-resonant Raman cooling is performed (red diamonds, y-axis offset for display).  The blue circles correspond to 10 ms of near-resonant Raman cooling on the red sideband with an observed $\bar{n}$ = 0.15(6).  Data points in each data set were taken randomly and interleaved.  Error bars are given by binomial statistics.  The slight shift in carrier frequency between the traces is due to magnetic field drift changing the Zeeman splitting of the S$_{1/2}$ state.}
\end{figure*}

The magnetic field that provides the Zeeman-splitting in the apparatus is produced by a Helmholtz coil driven by a constant-current power supply.  Using a fluxgate magnetometer nearby and scaling the observed magnetometer fluctuation with the DC field at trap center as measured by the ion, we find that the magnetic field at trap center drifts by $\sim$12 mG.  The corresponding drift in the Zeeman splitting is $\sim$30 kHz, which is near the fitted width of the sidebands shown in Fig.~\ref{fig:FORTemp}.  Stabilizing this field can be done by using a servo loop with the fluxgate magnetometer.  This would allow for longer data collection runs and permit more reliable 4-point temperature measurements.

The sideband Rabi frequency can be increased with additional far-off-resonant laser power and less detuning from the S$_{1/2}$ $\rightarrow$ P$_{1/2}$ transition, each providing a linear increase in the Rabi frequency of the Raman transition.  Both are currently limited by the small island of stability of the far-off-resonant 493 nm laser source.  Replacing the laser diode in the ECDL with a higher power diode with a center wavelength closer to 493 nm will allow both higher power and improved tunability.

Increasing the experimental state detection contrast can be achieved by improving three contrast reduction mechanisms in the experiment: (1) the far-off-resonant pulse decoherence rate, (2) the protect stage, and (3) the shelving stage.

Since the current far-off-resonant Raman decoherence rate is likely dominated by residual thermal motion in the radial modes, the decoherence rate could be decreased by actively stabilizing the radial secular frequencies and applying Raman sideband cooling to the radial modes as well.  This would reduce decoherence of the far-off-resonant Raman pulses, increasing detection contrast.  Note that this would provide an additional contrast increase when combined with a higher sideband Rabi frequency.

To improve the protect stage, a Raman technique could again be used with the Raman pump beam and the 650 nm repump laser to selectively transfer population from the S$_{1/2}$ $m_j=+\frac{1}{2}$ Zeeman sublevel to the D$_{3/2}$ protected manifold, which has been demonstrated previously.  This would increase the fidelity of the protect stage from 50\% to $\sim$90\% \cite{Chuah2013}.  Stimulated Raman adiabatic passage (STIRAP) could also be used, which has been previously demonstrated in a similar system \cite{DeVoe2002}.  These techniques could also serve both as the motional state detection and following state readout technique, replacing the far-off-resonant Raman source and state readout technique used here.  Note that STIRAP between the two S$_{1/2}$ Zeeman sublevels using either the near- or far-off-resonant Raman beams, rather than the far-off-resonant 493 nm Raman laser, does not work due to the additional coupling of the $\pi$-polarized Raman probe beam to the P$_{1/2}$, $m_j=-\frac{1}{2}$ Zeeman sublevel.  This prevents a suitable adiabatic transfer state in the dressed atom picture \cite{Vitanov1999}.

To improve the shelving stage, a Raman or STIRAP technique could similarly be used with the 455 nm and 614 nm laser sources.  This has previously been demonstrated \cite{Lewty2013}, increasing the shelving contrast from (88\%-12\%) = 76\% to $\sim$95\%.  Although all mentioned improvements would require at most one additional far-off-resonant 650 nm laser to couple between the P$_{1/2}$ and D$_{3/2}$ manifolds, the laser linewidth and locking requirements to address these dipole-allowed transitions are significantly more relaxed than those required for the 1.76 $\mu$m quadrupole laser.  Combining all contrast-increasing solutions, we expect a spin-state detection contrast $>$90\%, a factor of 2 improvement over the current contrast.

\section{Outlook and Conclusion\label{sec:conclusion}}

We have realized motional ground state cooling in $^{138}$Ba$^+$ using Raman sideband cooling with the comparatively simple Zeeman structure of the S$_{1/2}$ state rather than a barium isotope with hyperfine structure.  Our near-resonant Raman sideband cooling method requires no additional lasers to those required for Doppler cooling, and we estimate that our final $\bar{n}\approx0.17$ using optical pumping fluorescence resulting from successful near-resonant Raman transitions.

Employing a second, far-off-resonant laser driving Raman $\pi$-pulses between the two Zeeman sublevels and using a spin-dependent shelving technique, we verify our final motional occupation number of $\bar{n} = 0.15(0.06)$, and that optimal near-resonant Raman sideband cooling frequency corresponds to the red sideband.  Along with previously demonstrated optical cooling of aluminum hydride (AlH$^+$) \cite{Lien2014}, these simple techniques for $^{138}$Ba$^+$ will allow efficient molecular QLS to be performed using a 2-ion crystal.  Molecular QLS represents an important step forward for molecular quantum technology and will enable ultrahigh precision spectroscopy for fundamental studies, such as the search for time-variation of the proton-electron mass ratio.

\begin{acknowledgements}
This work was supported by AFOSR Grant No. FA9550-13-1-0116, NSF Grant No. PHY-1404455, and NSF GRFP DGE-1324585
\end{acknowledgements}

\clearpage

\clearpage

\clearpage

\clearpage

\clearpage
\begin{figure*}[h]
\centering
\includegraphics[width=\textwidth,height=0.75\textheight,keepaspectratio]{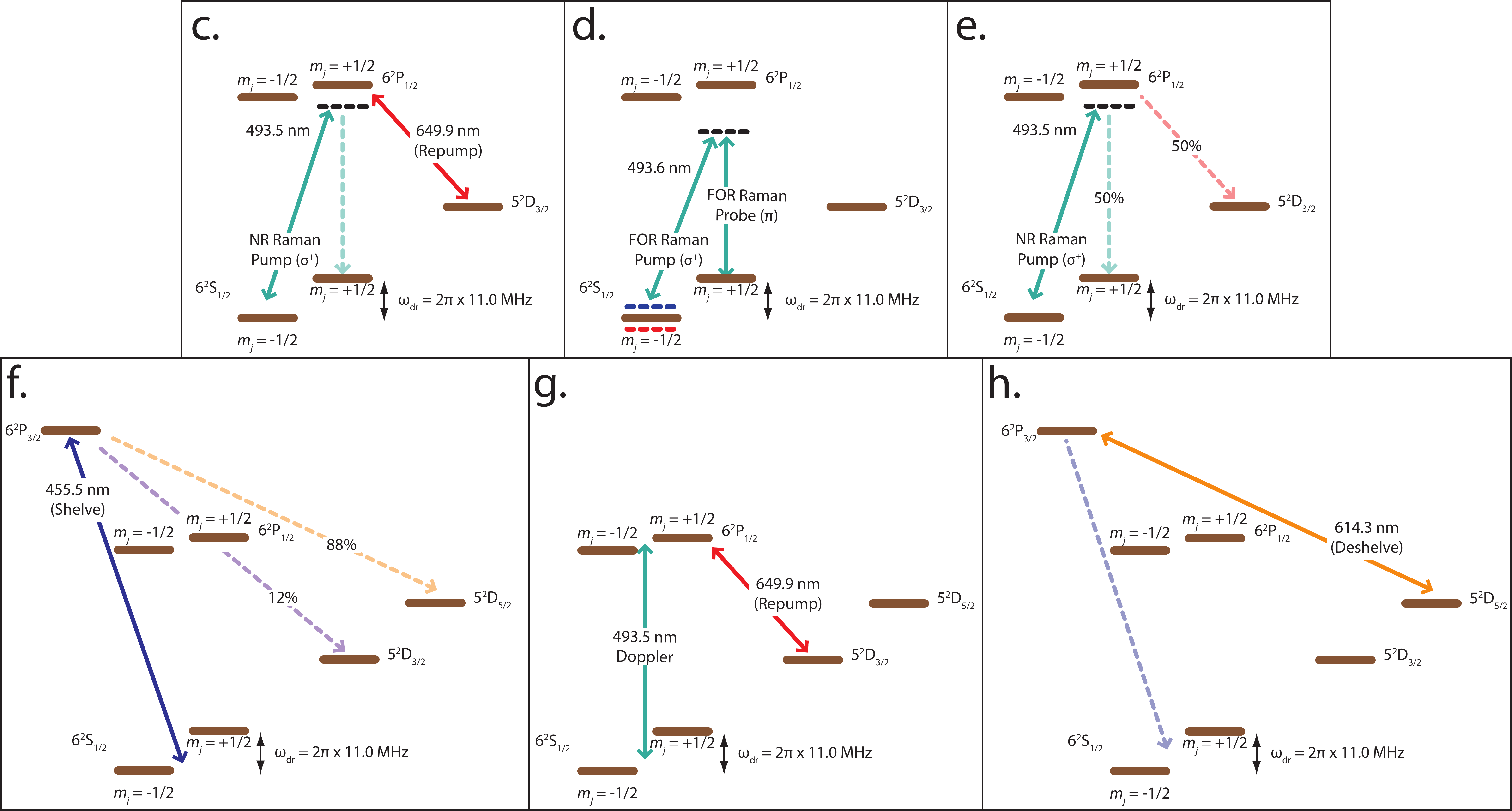}
\caption{\label{fig:FORRamanTiming}Stages of $^{138}$Ba$^+$ far-off-resonant Raman state detection that occur after near-resonant Raman cooling steps (a) and (b) in Fig.~\ref{fig:CWRamanTiming}.  Solid arrows show applied laser drives, and dashed arrows indicate dominant spontaneous emission channels with corresponding branching fractions.  Horizontal dashed lines show the red and blue motional sidebands of the driven Raman transition when relevant and the virtual state of the Raman transition (black), detuned by $\approx80$ MHz ($\approx59$ GHz) for the near-resonant (far-off-resonant) Raman.  (c) Optically pump to the $m_j=+\frac{1}{2}$ state.  (d) Far-off-resonant Raman $\pi$-pulse, conditional on motional state.  (e) Protect population in $m_j=-\frac{1}{2}$ by pumping to D$_{3/2}$.  (f) Shelve any population remaining in S$_{1/2}$ to D$_{5/2}$.  (g) Determination of D$_{3/2}$ versus D$_{5/2}$ ion state by observing fluorescence with the PMT.  (h) Deshelve any population in D$_{5/2}$ and additional Doppler cooling (not shown).}
\end{figure*}

\clearpage

\clearpage

\clearpage

\begin{appendix}
\clearpage
\section{Determining Optical Pumping Time and State Readout Fidelity\label{OPTAppendix}}

After near-resonant Raman sideband cooling, the population is optically pumped to the $m_j=+\frac{1}{2}$ state using the near-resonant Raman pump beam with the 650 nm repump beam on.  The repump beam ensures that population in the target $m_j=+\frac{1}{2}$ state is not pumped to the D$_{3/2}$ manifold by polarization impurities and magnetic field imperfections, allowing the optical pumping time to be arbitrarily long.  However, to minimize the effects of ion heating after the Raman sideband cooling on the final $\bar{n}$ measurement, this time should only be long enough to allow for maximum detection contrast.

To determine the optimum optical pumping time, the procedure described in Fig.~\ref{fig:FORRamanTiming} is performed.  Using the optimal protect time discussed in App.~\ref{PTAppendix}, the optical pumping is varied.  Both experiments allow measurement of the optical pumping time, but they also allow us to understand different state preparation fidelities in the experiment.  To prevent complications from the near-resonant Raman sideband cooling, this was left off for these data taken below.

\begin{figure*}[h]
\includegraphics[width=1.0\textwidth]{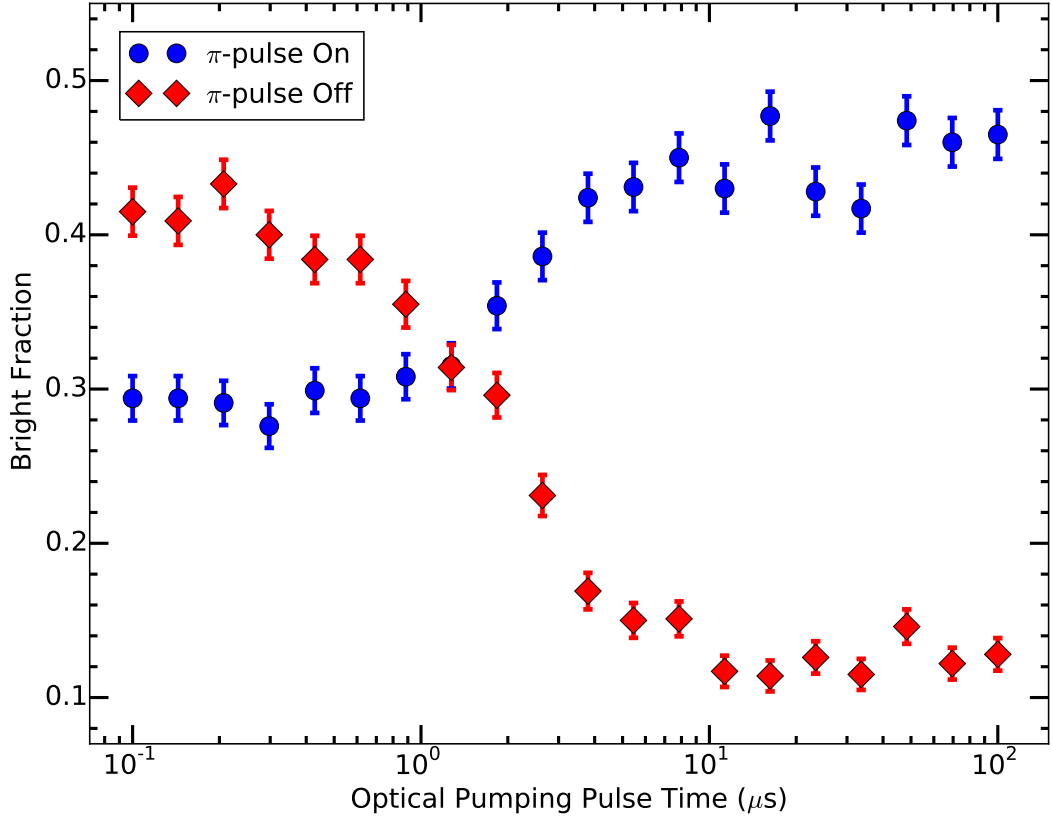}
\caption{\label{fig:OPTime} Determination of optimal optical pumping time for the step shown in Fig.~\ref{fig:FORRamanTiming}c.  Far-off-resonant $\pi$-pulse off (red diamonds).  Far-off-resonant $\pi$-pulse on (blue circles).  We see that the minimum optical pumping time for minimum (maximum) bright fraction with the far-off-resonant $\pi$-pulse off (on) is $\approx$10 $\mu$s.  Data points in each data set were taken randomly and interleaved.  Error bars are given by binomial statistics.}
\end{figure*}

Without the far-off-resonant $\pi$-pulse, we want to find the shortest optical pumping time that minimizes the bright fraction.  At long optical pumping times, we expect that the population will always be in $m_j=+\frac{1}{2}$ with a resulting 12\% bright fraction due to the branching ratio from P$_{3/2}$ to D$_{3/2}$.  At short times, the population should have equal probability of being in either Zeeman sublevel.  The protect stage then places the population starting in $m_j=-\frac{1}{2}$ to the protected D$_{3/2}$ state with 50\% efficiency ($\Gamma_{650}/\Gamma_{493}\approx\frac{1}{3}$); after the protect stage, 75\% of the population is in the S$_{1/2}$, $m_j=+\frac{1}{2}$ state with 25\% in the protected D$_{3/2}$ state.  After shelving and accounting for the 12\% branching ratio from P$_{3/2}$ $\rightarrow$ D$_{3/2}$, the ion should be bright with probability $\left(25\%+12\%\times75\%\right)=34\%$ \cite{Koerber2003}.  Experimentally at long optical pumping time, we observe the population saturate at the expected bright fraction of 12\% with 10 $\mu$s of optical pumping.  At very short optical pumping times, we observe a higher-than-expected initial bright fraction of $\approx$42\% (red diamonds in Fig.~\ref{fig:OPTime}).

With the far-off-resonant $\pi$-pulse, we want to verify the optical pumping time determined above by finding the shortest optical pumping time that maximizes the bright fraction.  At long optical pumping times, the ion is optically pumped to $m_j=+\frac{1}{2}$ before the $\pi$-pulse, and is maximally bright.  For this experiment, the maximum bright fraction of (44\%+12\%) = 56\% was not reached due to fast decoherence of the far-off-resonant Raman carrier without Raman sideband cooling.  However, the maximum brigh fraction occurs after 10 $\mu$s of optical pumping.  At very short optical pumping times, we expect that the population would be independent of optical pumping time, and the population should be 34\% bright.  We note a decrease from this expected value by $\approx$5\% (blue circles in Fig.~\ref{fig:OPTime}), which correlates to the bright fraction excess without the far-off-resonant $\pi$-pulse above.

With our Zeeman S$_{1/2}$ splitting of $\omega_\text{dr}\sim\frac{1}{2}\Gamma$ and the smaller Zeeman splitting of the P$_{1/2}$ state, we expect some degree of optical pumping by Doppler cooling.  When the $\pi$-polarized Doppler cooling frequency is fixed at $\approx\Gamma/2$ based on ion fluorescence, the $m_j=+\frac{1}{2}$ transition is closer to resonance than the $m_j=-\frac{1}{2}$ transition.  In a steady-state regime, this should result in a population excess in $m_j=-\frac{1}{2}$, which we observe in the procedure without the far-off-resonant $\pi$-pulse.  With the far-off-resonant Raman $\pi$-pulse, the populations in each $S_{1/2}$ Zeeman sublevel exchange, resulting in the observed population deficiency using this procedure.

From both data set results, the shortest optical pumping time that allows for the maximum detection contrast is 10 $\mu$s.

\clearpage
\section{Determining Protect Time\label{PTAppendix}}

After the far-off-resonant pulse, the near-resonant Raman pump beam is turned on for a short duration ("protect" time) with the 650 nm repump beam off.  Since this process is also limited by the rate of off-resonantly scattered photons, we expect this protect time to be similar to the optical pumping time of 10 $\mu$s as discussed in App.~\ref{OPTAppendix}.

If this time is too short, insufficient photon scatters will result in reduced optical pumping.  If this time is too long, polarization impurities in the near-resonant Raman pump beam and magnetic field imperfections will cause undesired scattering events from the $m_f=+\frac{1}{2}$ state to the D$_{3/2}$ manifold.  The efficiency of this step is limited to 50\% by the branching ratios of the P$_{1/2}$ state to each S$_{1/2}$ Zeeman sublevel and the D$_{3/2}$ manifold.  Combined with the shelving efficiency of 88\%, the maximum spin-state detection contrast is 44\%.

\begin{figure*}[h]
\includegraphics[width=1.0\textwidth]{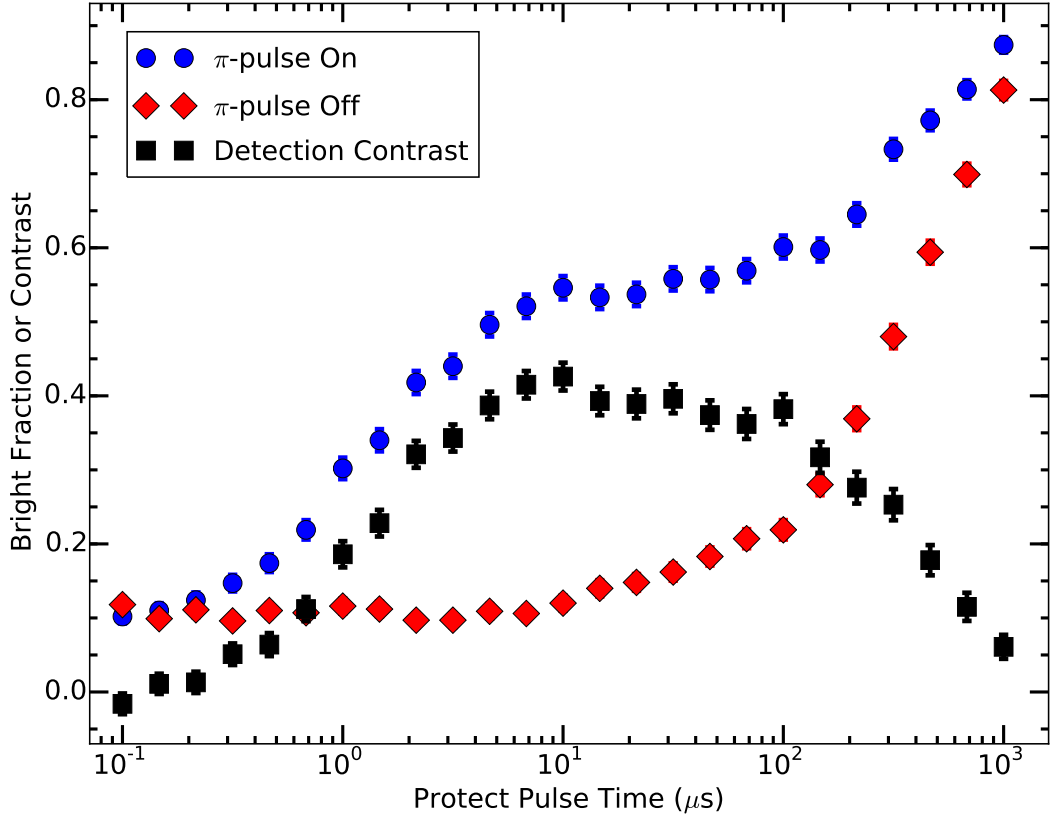}
\caption{\label{fig:PTTime} Determination of optimal protect pulse time for the step shown in Fig.~\ref{fig:FORRamanTiming}e.  The detection contrast (black squares) is calculated as the difference between the experiment without the far-off-resonant Raman $\pi$-pulse (red diamonds) and with the far-off-resonant $\pi$-pulse (blue circles).  The detection contrast is maximum at a protect pulse time of $\approx$10 $\mu$s as expected from the minimum optical pumping time determined in App.~\ref{OPTAppendix}.  Data points in each data set were taken randomly and interleaved.  Error bars are given by binomial statistics.}
\end{figure*}

To determine the optimal protect time, the procedure described in Fig.~\ref{fig:FORRamanTiming} was performed as the protect time was scanned without and with the far-off-resonant $\pi$-pulse (Fig.~\ref{fig:PTTime}).  Without the far-off-resonant $\pi$-pulse, the resultant bright ion fraction is only from polarization impurities and magnetic field imperfections (red diamonds).  With the far-off-resonant Raman $\pi$-pulse, the resultant bright ion fraction is from both a successful Rabi $\pi$-pulse and background effects (blue circles).  Subtracting the two signals yields the detection contrast, which is used to obtain the optical protect time of 10 $\mu$s, as expected from the optimal optical pumping time.  The maximum experimental detection contrast is 43(2)\% (Black squares), consistent with the maximum theoretical of 44\%.

\end{appendix}
\end{document}